\documentclass[aps,showpacs,nofootinbib,twocolumn]{revtex4}
\usepackage{amsmath}
\usepackage{mathtools}
\usepackage{epsfig,amssymb,amsmath}
\usepackage{subfigure}
\usepackage{CJK}
\usepackage[english]{babel}
\newtheorem{theorem}{Uncertainty relation}
\begin{document}
\title{Uncertainty equalities and uncertainty relation in weak measurement}
\author{Qiu-Cheng Song$^{1}$}
\author{Cong-Feng Qiao$^{1,2}$\footnote{Corresponding author, qiaocf@ucas.ac.cn}}
\affiliation{$^1$School of Physics, University of Chinese Academy of Sciences - YuQuan Road 19A, Beijing 100049, China\\
$^2$CAS Center for Excellence in Particle Physics, Beijing 100049, China}

\begin{abstract}
 Uncertainty principle is one of the fundamental principles of quantum mechanics. In this work, we derive two uncertainty equalities, which hold for all pairs of incompatible observables. We also obtain an uncertainty relation in weak measurement which captures the limitation on the preparation of  pre- and post-selected ensemble and hold for two non-Hermitian operators corresponding to two non-commuting observables.

\end{abstract}
\maketitle

\section{Introduction}
Uncertainty principle is one of the basic tenets of quantum mechanics. The initial spirit of uncertainty principle was postulated by Heisenberg \cite{heis}. Kennard first mathematically derived the Heisenberg uncertainty relation \cite{Kennard}. The most famous and popular form is the Heisenberg-Robertson uncertainty relation \cite{Robertson}
\begin{eqnarray}\label{ineqr1}
\Delta A^2\Delta B^2\geq
|\frac{1}{2}\langle\psi|[A,B]|\psi\rangle|^2\ ,
\end{eqnarray}
for any observables $A$, $B$, and any state $|\psi\rangle$, where the
variance of an observable $X$ in state $|\psi\rangle$ is defined as
$\Delta X^2=\langle\psi|X^2|\psi\rangle-\langle\psi|X|\psi\rangle^2$ and
the commutator is defined as $[A,B]=AB-BA$. A stronger extension of the Heisenberg-Robertson uncertainty relation (\ref{ineqr1}) was made by Schr\"odinger \cite{schroedinger}, which is generally formulated as
\begin{eqnarray}\label{ineqr2}
\Delta A^2\Delta B^2\geq\left|\frac 12 \langle[A,B]\rangle\right|^2
+\left|\frac{1}{2}\langle\{A,B\}\rangle - \langle A\rangle\langle B\rangle\right|^2 ,
\end{eqnarray}
where the anticommutator is defined as $\{A,B\}=AB+BA$, and $\langle X\rangle$ is defined as the expectation value $\langle\psi| X|\psi\rangle$ for any operator $X$ with respect to the normalized state $|\psi\rangle$.

However, the above two uncertainty relations have the problem that they may be trivial even when $A$ and $B$ are incompatible on the state $|\psi\rangle$. In order to correct this problem, Maccone and Pati \cite{mp} presented two stronger
uncertainty relations based on the sum of variances. The first one reads
\begin{eqnarray}\label{ineqr3}
\Delta A^2 + \Delta B^2\geq
\pm i|\langle[A,B]\rangle|+|\langle\psi|A \pm iB|\psi^\perp\rangle|^2\ ,
\end{eqnarray}
which is valid for arbitrary states $|\psi^\perp\rangle$ orthogonal to the state of the system state $|\psi\rangle$, where the sign should be chosen so that $\pm i\langle[A,B]\rangle$ (a real quantity) is positive. The second uncertainty relation is
\begin{eqnarray}\label{ineqr4}
\Delta A^2 + \Delta B^2\geq
\frac 12|\langle\psi^\perp_{A+B}|A+B|\psi\rangle|^2\ .
\end{eqnarray}
Here $|\psi^\perp_{A+B}\rangle\propto(A+B-\langle A + B\rangle)|\psi\rangle$ is a state orthogonal to $|\psi\rangle$. Maccone and Pati also derived an amended Heisenberg-Robertson uncertainty relation
\begin{eqnarray}\label{ineqr6}
\Delta A\Delta B\geq\frac{\pm i\frac 12\langle[A,B]\rangle}
{1-\frac 12|\langle\psi|\frac{A}{\Delta A}\pm i\frac {B}{\Delta B}|\psi^\perp\rangle|^2}\ ,
\end{eqnarray}
which is stronger than the Heisenberg-Robertson uncertainty relation (\ref{ineqr1}).

Recently, two stronger Schr\"odinger-like uncertainty relations have been proved which go beyond the
Maccone and Pati's uncertainty relation \cite{schlike}. The new relations provide stronger bounds whenever the observables
are incompatible on the state $|\psi\rangle$. The first uncertainty relation is
\begin{align}\label{ineqa}
\Delta A^2 + \Delta B^2\geq&
|\langle[A,B]\rangle + \langle\{A,B\}\rangle-2\langle A\rangle\langle B\rangle|\notag\\
&+|\langle\psi|A-e^{i\alpha}B|\psi^\perp\rangle|^2\ ,
\end{align}
which is valid for arbitrary states $|\psi^\perp\rangle$ orthogonal to the state of the system $|\psi\rangle$ and stronger than the Maccone and Pati's uncertainty relation (\ref{ineqr3}). Here $\alpha$ is a real constant, if $\langle \{A,B\} \rangle-2\langle A\rangle\langle B\rangle >0$, then  $\alpha = \arctan \tfrac{-i\langle[A,B]\rangle} {\langle\{A,B\} \rangle- 2\langle A\rangle\langle B\rangle}$; if $\langle\{A,B\}\rangle-2\langle A\rangle \langle B\rangle<0$, then $\alpha = \pi+\arctan \tfrac{-i\langle[A,B] \rangle} {\langle\{A,B\}\rangle-2\langle A\rangle\langle B\rangle}$; and while
$\langle\{A,B\} \rangle-2\langle A\rangle\langle B\rangle=0$, the relation (\ref{ineqa}) will reduce to (\ref{ineqr3})\ .
The second uncertainty relation is
\begin{eqnarray}\label{ineqb}
\Delta A^2\Delta B^2\geq\frac{\left|\tfrac 12\langle[A,B]\rangle\right|^2
+ \left|\tfrac{1}{2}\langle\{A,B\}\rangle-\langle A\rangle\langle B\rangle\right|^2}
{(1-\tfrac 12|\langle\psi|\tfrac{A} {\Delta A}-e^{i\alpha}\tfrac {B}{\Delta B}|\psi^\perp\rangle|^2)^2}\
\end{eqnarray}
which is stronger than the Schr\"odinger uncertainty relation (\ref{ineqr2}).

However, these new state dependent uncertainty relations have some problem \cite{bannur}, but some state independent uncertainty relations \cite{li,hang} immune from the drawback. Maccone and Pati's uncertainty relations \cite{mp} still are very important and have some generalizations. Two variance-based uncertainty equalities have been proved recently by Yao et al. \cite{sun} on the trend of stronger uncertainty relations \cite{mp}, for all pairs of incompatible observables $A$ and $B$. Meanwhile, two uncertainty relations in weak measurement were introduced by Pati et al. \cite{patiwu} for variances of two non-Hermitian operators corresponding to two non-commuting observables.

In this work we derive and proof two uncertainty equalities, which hold for all pairs of incompatible observables $A$ and $B$. We also give an uncertainty relation in weak measurement for two non-Hermitian operators corresponding to two non-commuting observables.

\section{Uncertainty equalities}

In this section, we construct and prove two uncertainty equalities which imply the uncertainty inequalities (\ref{ineqa}) and (\ref{ineqb}).
\begin{theorem}
\begin{align}\label{eqa}
\Delta A^2 + \Delta B^2=&
|\langle[A,B]\rangle + \langle\{A,B\}\rangle-2\langle A\rangle\langle B\rangle|\notag\\
&+\sum^{d-1}_{n=1}|\langle\psi|A-e^{i\alpha}B|\psi_n^\perp\rangle|^2\ ,
\end{align}
where $\{|\psi\rangle,|\psi^\perp_{n}\rangle_{n=1}^{d-1}\}$
comprise an orthonormal complete basis in the $d$-dimensional Hilbert space.
\end{theorem}
\emph{Proof}: To prove our uncertainty relation,
let us define the operator $\Pi=I-|\psi\rangle\langle\psi|$, $\bar{A}=A-\langle A\rangle I$, $\bar{B}=B-\langle B\rangle I $ and the state $|\phi\rangle=(\bar{A}-e^{i\tau}\bar{B})|\psi\rangle$, we have
\begin{align}\label{eqa1}
\langle\phi|\Pi|\phi\rangle
=&\langle\psi|(\bar{A}-e^{-i\tau}\bar{B})|(I-|\psi\rangle\langle\psi|)|(\bar{A}-e^{i\tau}\bar{B})|\psi\rangle\notag\\
=&\langle\psi|(\bar{A}-e^{-i\tau}\bar{B})(\bar{A}-e^{i\tau}\bar{B})|\psi\rangle\notag\\
=&\Delta A^2+\Delta B^2-2\text{Re}(e^{i\tau}\langle\psi| \bar{A}\bar{B}|\psi\rangle).
\end{align}
There exists $\tau=-\alpha$, so that $e^{i\tau}\langle\psi| \bar{A}\bar{B}| \psi\rangle$ is real, and it can be written as $|\langle\psi| \bar{A}\bar{B}| \psi\rangle|$, we obtain
\begin{align}\label{eqa2}
&\langle\psi|(\bar{A}-e^{i\alpha}\bar{B})|\Pi|(\bar{A}-e^{-i\alpha}\bar{B})|\psi\rangle\notag\\
=&\Delta A^2+\Delta B^2-2|\langle\psi|\bar{A}\bar{B}|\psi\rangle|\notag\\
=&\Delta A^2+\Delta B^2-|\langle[A,B]\rangle + \langle\{A,B\}\rangle-2\langle A\rangle\langle B\rangle|,
\end{align}
Since $\Pi$ is the orthogonal complement to $|\psi\rangle\langle\psi|$,
we can choose an arbitrary orthogonal decomposition of the projector $\Pi$
\begin{align}\label{eqa3}
\Pi=\sum^{d-1}_{n=1}|\psi^{\perp}_{n}\rangle\langle\psi^{\perp}_{n}|,
\end{align}
where $\{|\psi\rangle,|\psi^\perp_{n}\rangle_{n=1}^{d-1}\}$
comprise an orthonormal complete basis in the $d$-dimensional Hilbert space. Whence, Eq. (\ref{eqa2}) can be rewritten as
\begin{align}\label{eqa4}
&\sum^{d-1}_{n=1}|\langle\psi|(\bar{A}-e^{i\alpha}\bar{B})|\psi^{\perp}_{n}\rangle|^2\notag\\
=&\sum^{d-1}_{n=1}|\langle\psi|A-e^{i\alpha}B|\psi^{\perp}_{n}\rangle|^2\notag\\
=&\Delta A^2+\Delta B^2-|\langle[A,B]\rangle + \langle\{A,B\}\rangle-2\langle A\rangle\langle B\rangle|,
\end{align}
which is equivalent to (\ref{eqa}).

\begin{theorem}
\begin{eqnarray}\label{eqb}
\Delta A^2\Delta B^2=\frac{\left|\tfrac 12\langle[A,B]\rangle\right|^2
+ \left|\tfrac{1}{2}\langle\{A,B\}\rangle-\langle A\rangle\langle B\rangle\right|^2}
{(1-\tfrac 12\sum^{d-1}_{n=1}|\langle\psi|\tfrac{A} {\Delta A}-e^{i\alpha}\tfrac {B}{\Delta B}|\psi_n^\perp\rangle|^2)^2}\ ,
\end{eqnarray}
where $\{|\psi\rangle,|\psi^\perp_{n}\rangle_{n=1}^{d-1}\}$
comprise an orthonormal complete basis in the $d$-dimensional Hilbert space.
\end{theorem}

\emph{Proof}: To prove our uncertainty equality, let us define  the operator $\Pi=I-|\psi\rangle\langle\psi|$, $\bar{A}=A-\langle A\rangle I$, $\bar{B}=B-\langle B\rangle I$ and the unnormalized  state $|\phi\rangle=(\tfrac{\bar{A}} {\Delta A}-e^{i\tau}\tfrac {\bar{B}}{\Delta B})|\psi\rangle$, we have
\begin{align}\label{eqb1}
&\langle\phi|\Pi|\phi\rangle\notag\\
=&\langle\psi|(\tfrac{\bar{A}} {\Delta A}-e^{-i\tau}\tfrac {\bar{B}}{\Delta B})|(I-|\psi\rangle\langle\psi|)|(\tfrac{\bar{A}} {\Delta A}-e^{i\tau}\tfrac {\bar{B}}{\Delta B})|\psi\rangle\notag\\
=&\langle\psi|(\tfrac{\bar{A}} {\Delta A}-e^{-i\tau}\tfrac {\bar{B}}{\Delta B})(\tfrac{\bar{A}} {\Delta A}-e^{i\tau}\tfrac {\bar{B}}{\Delta B})|\psi\rangle\notag\\
=&2-2\frac{\text{Re}(e^{i\tau}\langle\psi| \bar{A}\bar{B}|\psi\rangle)}{\Delta A\Delta B},
\end{align}
There exists $\tau=-\alpha$, so that $e^{i\tau}\langle\psi| \bar{A}\bar{B}| \psi\rangle$ is real, and it can be written as $|\langle\psi| \bar{A}\bar{B}| \psi\rangle|$, we obtain
\begin{align}\label{eqb2}
&\langle\psi|(\tfrac{\bar{A}} {\Delta A}-e^{i\alpha}\tfrac {\bar{B}}{\Delta B})|\Pi|(\tfrac{\bar{A}} {\Delta A}-e^{-i\alpha}\tfrac {\bar{B}}{\Delta B})|\psi\rangle\notag\\
=&2-2\frac{|\langle\psi| \bar{A}\bar{B}|\psi\rangle|}{\Delta A\Delta B},
\end{align}
Similarly, we choose the projector $\Pi$
\begin{align}\label{eqa3}
\Pi=\sum^{d-1}_{n=1}|\psi^{\perp}_{n}\rangle\langle\psi^{\perp}_{n}|.
\end{align}
Then  Eq. (\ref{eqb2}) can be rewritten as
\begin{align}\label{eqb4}
&\sum^{d-1}_{n=1}|\langle\psi|(\tfrac{\bar{A}} {\Delta A}-e^{i\alpha}\tfrac {\bar{B}}{\Delta B})|\psi^{\perp}_{n}\rangle|^2\notag\\
=&\sum^{d-1}_{n=1}|\langle\psi|\tfrac{A} {\Delta A}-e^{i\alpha}\tfrac {B}{\Delta B}|\psi^{\perp}_{n}\rangle|^2\notag\\
=&2-2\frac{|\frac{1}{2}\langle[A,B]\rangle + \frac{1}{2}\langle\{A,B\}\rangle-\langle A\rangle\langle B\rangle|}{\Delta A\Delta B},
\end{align}
which is equivalent to (\ref{eqb}).

 The two uncertainty equalities (\ref{eqa}) and (\ref{eqb}) are valid for all pairs of incompatible observables. If we retain only one term associated with $|\psi^{\perp}\rangle\in\{|\psi^\perp_{n}\rangle_{n=1}^{d-1}\}$ in the summation and
discard the others, the uncertainty equalities (\ref{eqa}) and (\ref{eqb}) reduce to the uncertainty relations (\ref{ineqa}) and (\ref{ineqb}), respectively.

\section{Uncertainty relation in weak measurement}

First introduced by Aharonov, Albert, and Vaidman \cite{Aharonov},
weak values are complex numbers that one can define the weak value of $A$ using two states: an initial state $|\psi\rangle$, called the preselection, and a final state $|\varphi\rangle$, called the postselection. the weak value of $A$ has the form
\begin{align}
\langle A\rangle_w=\frac{\langle\varphi|A|\psi\rangle}{\langle\varphi|\psi\rangle}.
\end{align}
For a givern preselected and post-selected ensemble, define the operator $A_w$ as
\begin{align}
A_w=\frac{\Pi_\varphi A}{p},
\end{align}
where $\Pi_\varphi=|\varphi\rangle\langle\varphi|$ and $p=|\langle\varphi|\psi\rangle|^2$. This has many properties please reference \cite{patiwu}.

Here, we construct an uncertainty relation in weak measurement for variances of two non-Hermitian operators $A_w$ and $B_w$  corresponding to two non-commuting observables $A$ and $B$. The uncertainty relation quantitatively express the impossibility of jointly sharp preparation of pre- and post-selected (PPS) quantum states $|\psi\rangle$ and $|\varphi\rangle$ for the weak measurement of incompatible observables.
\begin{theorem}
\begin{align}\label{weak1}
\Delta & A^2_w+\Delta B^2_w\geq
|\frac{1}{p}\langle\varphi|[A,B]|\varphi\rangle
+\frac{1}{p}\langle\varphi|\{A,B\}|\varphi\rangle\notag\\
&-2\langle A_w\rangle\langle B_w\rangle^\ast|
+\left|\langle\psi|A_w-e^{i\alpha} B_w|\psi^\perp\rangle\right|^2.
 \end{align}
which is valid for two non-Hermitian operators $A_w$ and $B_w$, where $p$ is equivalent to $|\langle\varphi|\psi\rangle|^2$.
\end{theorem}

\emph{Proof}: To  prove this relation we define the variance for any general (non-Hermitian) operator $X$
 in a state $|\psi\rangle$ which can be defined as \cite{Anandan,AKPati}
 \begin{align}
 \Delta X^2=\langle\psi|(X-\langle X\rangle)(X^\dagger-\langle X^\dagger\rangle)|\psi\rangle.
 \end{align}
 The variance of the non-Hermitian operation $A_w$ in the quantum $|\psi\rangle$ can be defined as
 \begin{align}
 \Delta A_w^2=\langle\psi|(A_w-\langle A_w\rangle)(A_w^\dagger-\langle A_w^\dagger\rangle)|\psi\rangle,
 \end{align}
where $\langle A_w\rangle=\langle\psi| A_w|\psi\rangle$
and $\langle A_w^\dagger\rangle=\langle\psi| A_w^\dagger|\psi\rangle=\langle A_w\rangle^\ast$,
 $\Delta A^2_w$ can also be expressed as
\begin{align}
 \Delta A_w^2=\langle\psi|A_wA_w^\dagger\rangle|\psi\rangle-\langle\psi|A_w|\psi\rangle\langle\psi|A_w^\dagger|\psi\rangle.
 \end{align}
 Similarly, for Hermitian operator $B$, we can define the operator
 \begin{align}
B_w=\frac{\Pi_\varphi B}{p}.
 \end{align}
Then, the uncertainty for $B_w$ can also be defined as
 \begin{align}
 \Delta B_w^2=\langle\psi|B_wB_w^\dagger\rangle|\psi\rangle-\langle\psi|B_w|\psi\rangle\langle\psi|B_w^\dagger|\psi\rangle.
 \end{align}
To prove our uncertainty relation in weak measurement, we introduce a general inequality
\begin{align}\label{weak2}
\|C^\dagger|\psi\rangle
  -e^{i\tau}D^\dagger|\psi\rangle
  +k(|\psi\rangle-|\bar{\psi}\rangle)\|^2 \geq 0,
 \end{align}
 where $C^\dagger\equiv A^\dagger_w-\langle A^\dagger_w\rangle$ and
 $D^\dagger\equiv B^\dagger_w-\langle B^\dagger_w\rangle$.
By expanding the square modulus, we have
\begin{align}\label{weak3}
\Delta A^2_w+\Delta B^2_w\geq-\lambda k^2-\beta k+\pi,
\end{align}
where $\lambda\equiv 2(1-\text{Re}[\langle\psi|\bar{\psi}\rangle])$, $\pi\equiv2\text{Re}[e^{i\tau}\langle\psi|CD^\dagger|\psi\rangle]$, and $\beta\equiv 2\text{Re}[\langle\psi|(-C+e^{-i\tau}D)|\bar{\psi}\rangle]$. We choose the value of $k$
that maximizes the right-hand-side of (\ref{weak3}), namely $k=-\beta/2\lambda$, we get
\begin{align}
\Delta A^2_w+\Delta B^2_w\geq\frac{\beta^2}{4\lambda}+\pi.
\end{align}
The above inequality can be rewritten as
\begin{align}
\Delta A^2_w+\Delta B^2_w\geq
&\frac{\text{Re}[\langle\psi|(-C+e^{-i\tau}D)|\bar{\psi}\rangle]^2}{2(1-\text{Re}[\langle\psi|\bar{\psi}\rangle])}\notag\\
&+2\text{Re}[e^{i\tau}\langle\psi|CD^\dagger|\psi\rangle]
\end{align}
Suppose $|\bar{\psi}\rangle = \cos \theta| \psi\rangle + e^{i\phi} \sin\theta |\psi^{\perp}\rangle$,
where $| \psi^{\perp}\rangle$ is orthogonal to
$|\psi\rangle$, by taking the limit $\theta\rightarrow 0$,
the state $|\bar{\psi}\rangle$ reduces to $|\psi\rangle$ and
then the above inequality can be reexpressed as
\begin{align}\label{weak4}
\Delta A^2_w+\Delta B^2_w\geq
&\text{Re}[e^{i\phi}\langle\psi|(-A_w+e^{-i\tau}B_w)|\psi^\perp\rangle]^2\notag\\
&+2\text{Re}[e^{i\tau}\langle\psi|CD^\dagger|\psi\rangle],
\end{align}
there exists $\tau=-\alpha$ so that $e^{i\tau}\langle\psi|CD^\dagger|\psi\rangle$ is real,
 and it can be written as $|\langle\psi|CD^\dagger|\psi\rangle|$ ,
 and then the second term becomes $\{\text{Re}[e^{i\phi}\langle\psi|-A_w+e^{i\alpha}B_w|\psi^\perp\rangle]\}^2$,
  we can choose $\phi$ so that this term in square brackets is real,
  so that this term can be expressed as $|\langle\psi|A_w-e^{i\alpha}B_w|\psi^\perp\rangle|^2 $.
  Whence, inequality (\ref{weak4})  becomes
\begin{align}
\Delta A^2_w+\Delta B^2_w\geq &
|\langle\psi|A_w-e^{i\alpha}B_w|\psi^\perp\rangle|^2\notag\\
&+2|\langle\psi|CD^\dagger|\psi\rangle|.
\end{align}
The last term can be rewritten as
\begin{align}\label{weak5}
2|\langle CD^\dagger\rangle|
=|\langle CD^\dagger+DC^\dagger+CD^\dagger-DC^\dagger\rangle|,
\end{align}
where
\begin{align}\label{weak6}
&\langle CD^\dagger+DC^\dagger\rangle\notag\\
=&\frac{1}{p}\langle\varphi|\{A,B\}|\varphi\rangle
-\langle A_w\rangle\langle B_w\rangle^\ast-\langle A_w\rangle^\ast\langle B_w\rangle
\end{align}
and
\begin{align}\label{weak7}
&\langle CD^\dagger-DC^\dagger\rangle\notag\\
=&\frac{1}{p}\langle\varphi|[A,B]|\varphi\rangle
-\langle A_w\rangle\langle B_w\rangle^\ast
+\langle A_w\rangle^\ast\langle B_w\rangle.
\end{align}
We combine Eqs. (\ref{weak6}) and (\ref{weak7}), Eq. (\ref{weak5}) becomes
\begin{align}\label{weak8}
&2|\langle CD^\dagger\rangle|\notag\\
=&\left|\frac{1}{p}\langle\varphi|[A,B]|\varphi\rangle+\frac{1}{p}\langle\varphi|\{A,B\}|\varphi\rangle
-2\langle A_w\rangle\langle B_w\rangle^\ast\right|.
\end{align}
Combining Eqs. (\ref{weak5}) and (\ref{weak8}), we obtain the uncertainty relation (\ref{weak1}).

\section{Conclusions}

In this work, we derived two new uncertainty equalities for sum and product of variances of a pair of incompatible observables, which hold for all pairs of incompatible observables $A$ and $B$. In fact, one can obtain a series of inequalities  by retaining $1$ to $d-2$ terms within the set $\{|\psi^\perp_{n}\rangle_{n=1}^{d-1}\}$. We also derived an uncertainty relation in weak measurement for two non-Hermitian operators $A_w$ and $B_w$ corresponding to two non-commuting observables $A$ and $B$.  The uncertainty relation quantitatively expresses the impossibility of jointly sharp preparation of PPS quantum states $|\psi\rangle$ and $|\varphi\rangle$ for measuring incompatible observables during the weak measurement.

\vspace{.1cm}
{\bf Acknowledgments}
We are grateful to Junli Li for discussion. This work was supported in part by National Key Basic Research Program of China under the grant 2015CB856700, and by the National Natural Science Foundation of China(NSFC) under the grants 11175249 and 11375200.


\end{document}